\documentclass[
preprint,% this is the default
author-year,%choose either numerical or author-year
showpacs,
showkeys,
endfloats*
]{jasatex}

\usepackage{graphicx}% Include figure files
\usepackage{dcolumn}% Align table columns on decimal point
\usepackage{amsmath}
\usepackage{ amsfonts}% popular packages from the American Mathematical Society
\usepackage{bm}% Bold Math package
\usepackage[letterpaper,top=1in,bottom=1in,hmargin={1in,1in}]{geometry}

\usepackage{fancyhdr}
\fancyheadoffset[L,R]{\marginparsep}

\begin{document}

%\preprint{AIP/123-QED}

%\input{NS01_main}

\title[Noise Suppression for speech during fMRI]{Algorithm to suppress scanner noise in recorded speech during functional magnetic resonance imaging}

%{An algorithm for suppressing scanner noise and recording speech from a single-channel microphone during fMRI}
%A speech recording setup for fMRI with online reduction of scanner noise}% Force line breaks with \\
 
\author{Satrajit S. Ghosh\\satra@mit.edu}
\affiliation{%
McGovern Institute for Brain Research, Massachusetts Institute of Technology,  Cambridge, MA} 

\date{\today}% It is always \today, today,

\begin{abstract}
The high-intensity, repetitive noise associated with functional magnetic resonance imaging hinders on-line monitoring of subjects' speech and/or recording speech signals suitable for off-line analysis. The proposed algorithm enhances the speech signal by suppressing the scanner noise in the signal recorded by a single-channel microphone. Significant increases in signal-to-noise ratio are achieved using an adaptive filter that combines time and frequency domain elements. In addition to providing a recording suitable for speech analysis, such a real-time system provides an alternative means (to, e.g., the ``panic ball'') for communication between the patient and the operator during image acquisition.

%Valid PACS numbers may be entered using the \verb+\pacs{#1}+ command.
\end{abstract}

%Use showkeys class option if keyword
\maketitle

\section{I\lowercase{ntroduction}}
During a functional magnetic resonance imaging (fMRI) experiment, loud noise generated by the gradient coils of the scanner typically accompanies the acquisition of brain images \citep{ravi.acoustic.00}. The intensity of this noise can vary between 100 and 120 dBA SPL and the energy content is mostly concentrated in frequencies below 3 KHz, which is also the frequency range most relevant to speech signals. Therefore, such noise is particularly detrimental to recording speech in the scanner. Being able to record speech is not only important for a variety of speech and language related studies, but also provides a natural mechanism for a person in the scanner to communicate with the operator in the control room. Prior efforts to reduce such noise utilized frequency-domain spectral subtraction or time-domain template subtraction approaches (Table~\ref{tab:table1}). 

\begin{table}
\caption{\label{tab:table1}Different software based approaches used in the past for cancellation of fMRI noise. (T-Domain : time domain, F-Domain : frequency domain, NRT : near real-time)}
\begin{ruledtabular}
\begin{tabular}{lcccc}
Reference	& T-Domain	& F-Domain	& Adaptive & NRT\\
\hline
\citet{nell.automated.03}	&	& x	& & \\
\citet{jung.extraction.05}	& x	& &	x &  \\
\citet{cusa.automated.05}	& x	& & & \\	
Current Proposal	& x	& x	& x & x\\
\end{tabular}
\end{ruledtabular}
\end{table}

The spectral subtraction approach \citep{nell.automated.03} removed stationary noise from measurement by subtracting a noise magnitude spectrum estimate from successive short-time spectral estimates of the overall signal, and inverting using the overall signal phase information (Boll, 1979). The effectiveness of this approach relies on separation between the spectral properties of the speech signal and the noise source and is limited in this scenario where the scanner noise spectrum overlaps the speech spectrum. Template-based subtraction \citep{cusa.automated.05,jung.extraction.05} eliminates characteristic noise signals by simple subtraction of a time-domain template from the overall measurement in noise-correlated time frames. Such an approach requires high temporal sampling rate for accurate template matching. Both approaches in their simplest form assume noise properties are constant. The algorithm proposed here combines time and frequency domain elements into an adaptive approach implemented as a real-time software program that processes the acoustic signal acquired using off the shelf hardware components. 

The details of the algorithm are presented, followed by the results of simulations using synthetic data and of using the system during an fMRI experiment.

\section{B\lowercase{ackground}}
The signal, $y(t)$, recorded by a microphone placed near the mouth of the subject in the scanner comprises three different components: (1) the voice signal, $v(t)$, if present; (2) the scanner gradient noise, $g(t)$; and (3) other extraneous noise sources, $n(t)$, present in the environment (e.g. the helium pump, breathing). 

\begin{equation}
y(t) = v(t) + g(t) + n(t)
\label{eq:micsignal}
\end{equation}

\begin{figure}
\includegraphics[width=.9\columnwidth]{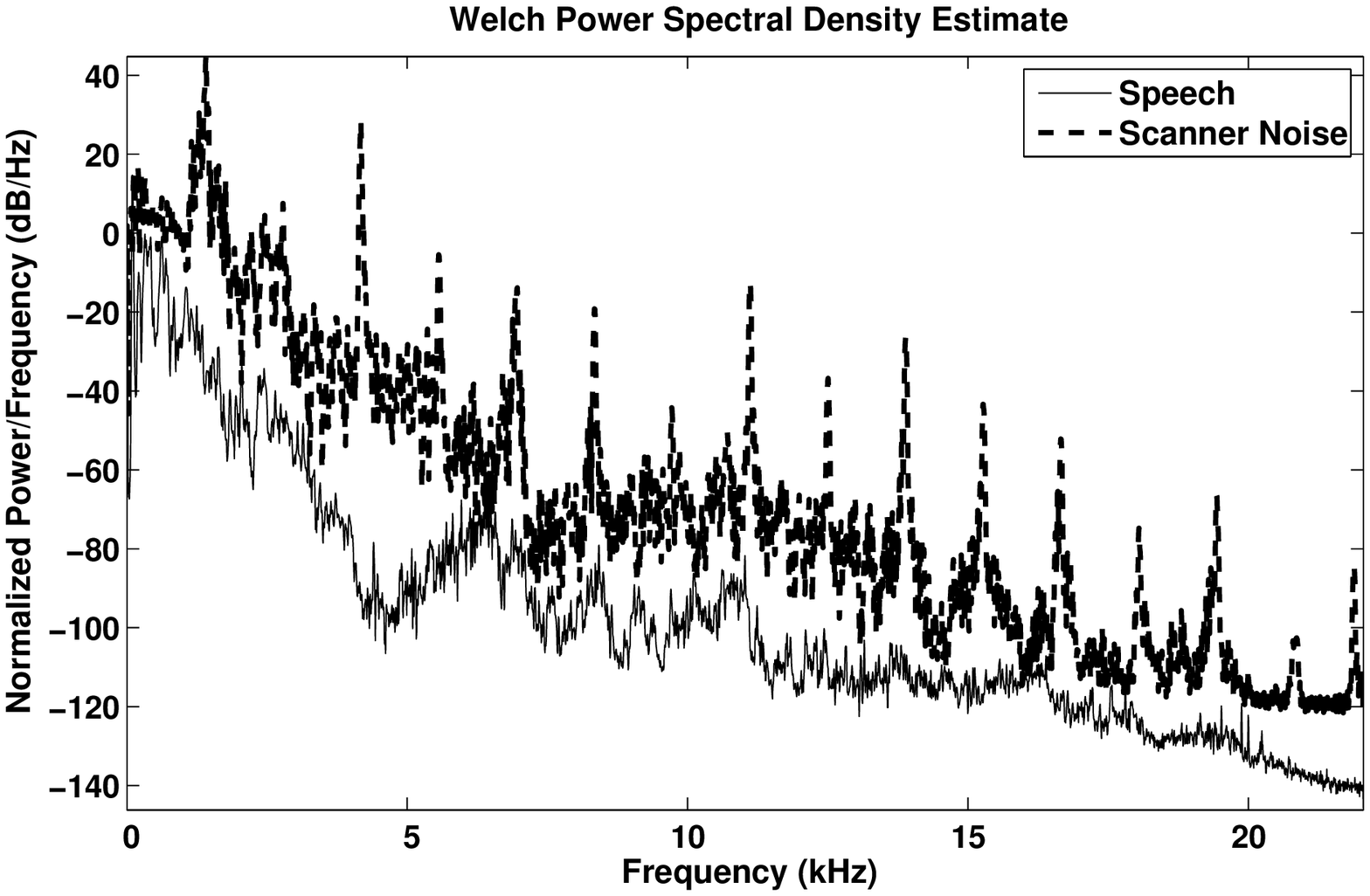}
\caption{Comparison of power spectral density of speech (black) and scanner noise (gray) during fMRI. The scanner noise was collected from a 3 Tesla Siemens Trio scanner. The speech signal was recorded from a male speaker. Power spectral density estimates were obtained using the Welch algorithm with a window of 100ms and an overlap of 80\%.}
\label{fig:spectraloverlap}
\end{figure}

The goal of the real-time algorithm is to estimate and continually refine $g(t)$ and to use the estimate in order to recover $v(t)$.  Most of the energy content of $g(t)$ is concentrated below 5kHz and the spectral components of $g(t)$ overlap $v(t)$ as shown in Figure~\ref{fig:spectraloverlap}.

Based on recorded scanner noise data, the algorithm assumes that \[|g(t)|>>|v(t)|>|n(t)|\] and that $g(t)$ is periodic. Currently, images during functional scanning are acquired in a planar manner, with a stack of two-dimensional (2-D) images making up a 3-D image volume. During an fMRI session, the role of gradient-switching is to select the plane from which to acquire the MR signal. It is this switching that generates the scanner ``noise'' and hence the periodicity of $g(t)$ is dependent on the time between the onset of two planar acquisitions. Thus, 

\begin{equation}
g(t) \approx g(t + \frac{T}{n})
\label{eq:periodicity}
\end{equation}

where, $T$ is the time taken to acquire $n$ 2-D images to create the 3-D volume. 

\section{A\lowercase{lgorithm}}
The algorithm has three stages. The first stage involves a time-domain estimation in order to initialize the noise template. In the second stage, the estimated template is matched to the signal. In the third and final stage, the template is subtracted from the matched segment. The template is updated if the matched segment does not contain speech. The second and third stages are repeated iteratively over the duration of recording.  Table~\ref{tab:table2} lists and describes the parameters and the signal vectors used in the algorithm, the details of which are described next. It is important to note that the algorithm operates sequentially on buffers of length $N$ and has a processing delay $\tau$ (typically less than 200ms) related to the buffer length and the estimated duration of the scanner noise template.

\begin{table}
\caption{\label{tab:table2}Algorithm parameters and signal vectors}
\begin{ruledtabular}
\begin{tabular}{ll}
Parameters & \\
\hline
$l_{est}$ & Estimated duration of noise template (s)\\
 & $l_{est} \approx T/n$\\
$\pm w$ & Variation in $l_{est}$ (s)\\
$s_r$ & sampling frequency (Hz)\\
$N$ & Framelength ($0.02s_r$ samples)\\
$\tau$ & Buffer length (samples)\\
 & $\tau = N + 2(l_{est}+w)s_r$\\
$\theta_{xcorr}$ & cross-correlation threshold for estimating \\
& template \\
$\theta_{corr}$ & correlation threshold for estimating \\
& signal match to template\\
$\alpha$ & Noise spectrum scaling parameter\\
$\theta_{RMS}$ & Template update threshold\\
$\gamma$ & Template update parameter\\
\hline
Signal vectors & \\
\hline
$\mathbf{\hat{g}}$ & Estimated noise template\\
$\mathbf{x_b}$ & Matched signal from buffer\\
$\mathbf{\hat{v}}$ & Noise suppressed signal\\
$\mathbf{x_{res}}$ & $\mathbf{x_b}-\mathbf{\hat{g}}$ \\
$\mathbf{\mathcal{F}\mathbf{x}}$ & Fourier transform of $\mathbf{x}$\\
$\mathbf{w}$ & Weighting function on the magnitude \\
& spectrum of the noise template\\
$\mathbf{d}$ & Digital filter (e.g., low pass filter at 5 KHz)\\
\end{tabular}
\end{ruledtabular}
\end{table}

\subsection{Step 1. Gradient noise template estimation}
The key to effective suppression is an accurate initial estimation of gradient noise template vector $\mathbf{g}$. A double sliding-window cross-correlation approach is used to estimate $\mathbf{g}$, in which the correlation between two adjacent windows of equal duration is calculated. This calculation is repeated over a short range of incremental window durations, since the periodicity can change by a few samples from slice to slice. The noise template, $\mathbf{\hat{g}}$, is set to the samples in the window for which this cross-correlation exceeds a pre-specified threshold $\theta_{xcorr}$. To maximize computational efficiency, the search for $\mathbf{\hat{g}}$ can be constrained by the periodicity of $g(t)$ ($l_{est} = T/n$; see Eq.~\ref{eq:periodicity}) and only window durations of $l_{est} \pm w$ are used.

\subsection{Step 2. Template matching }
Once $g(t)$ has been estimated, it can be correlated with samples in input audio buffer to determine a match. Computation time can be reduced by computing the correlations with lags between $N+1$ and $2N$, again leveraging the periodic property of $g(t)$. A template match occurs when the peak correlation over the span of lags exceeds a specified threshold $\theta_{corr}$. Similar to prior time-domain approaches, the noise template is then subtracted from the matched segment $\mathbf{x_b}$ to yield the residual $\mathbf{x_{res}}$. 

\subsection{Step 3. Template subtraction and update}
If $\mathbf{x_b}$ contained a speech signal and the template was a perfect match, then the residual $\mathbf{x_{res}}$ should only contain the speech signal. However, because of other noises $n(t)$ in the system, the residual may contain additional noise sources. To further enhance the speech signal, we perform a weighted frequency-domain subtraction (Eq.~\ref{eq:smagsub}) of the magnitudes of the spectral components. The weighting function $\mathbf{w}$ provides a mechanism to fine-tune the suppression. The estimated voice signal, $\mathbf{\hat{v}}$, is recovered by taking the inverse Fourier transform ($\mathcal{F}$) of this magnitude spectrum combined with the phase information from $\mathcal{F}\mathbf{x_{res}}$. An ideal digital filter ($\mathbf{d}$) may be incorporated at this stage to limit the bandwidth of the signal. If $\mathbf{\hat{v}}$ has minimal energy ($<\theta_{RMS}$), then most of the content of the buffer was likely scanner noise. This is then used to update the template $\mathbf{\hat{g}}$ (Eq.~\ref{eq:gupdate}). Equation~\ref{eq:gupdate} indicates that an increasing value of $\gamma$ will keep the estimates of $\mathbf{\hat{g}}$ similar over longer periods of time. Thus $\gamma$ can be used to control the similarity between successive updates of $\mathbf{\hat{g}}$. This dependence on prior estimates guards against periods of recording when the estimate is not updated due to the presence of voice in the signal.  

\begin{eqnarray}
\Gamma & = &\big[|\mathcal{F}\mathbf{x_{res}}|-\alpha\mathbf{w}\circ|\mathcal{F}\mathbf{\hat{g}}|\big]^+ \label{eq:smagsub}\\
\mathbf{\hat{v}} & = & \mathbb{R}\big(\mathcal{F}^{-1}\left(\mathbf{d}\circ\Gamma\circ e^{j\angle{\mathcal{F}\mathbf{x_{res}}}}\right)\big)\\
\mathbf{\hat{g}} & = & \gamma\mathbf{\hat{g}}+(1-\gamma)\mathbf{x_b} \text{\hbox{ }, when\hbox{ }} \text{RMS}(\mathbf{\hat{v}})<\theta_{RMS} \label{eq:gupdate}
\end{eqnarray}

In the equations above, $\circ$ denotes a Schur product and $[\cdot]^+$ indicates half-wave rectification.

\section{H\lowercase{ardware setup and data collection}}
All recordings were made with a Shure condenser microphone (model no: SM93). The microphone was placed inside a foam windscreen and was mounted on the headcoil a few cms from the subject's mouth. All ferromagnetic components of the microphone (primarily in the connectors) were stripped before using it inside the scanner room. The microphone cable to the supplied preamplifier was rerouted through RF filters mounted on the scanner filter panel. The preamplifier was connected to a MOTU audio device (model no: 828mkII), which supplied the microphone with the necessary phantom power. The setup did not introduce any artifacts in the acquired images or degrade it's quality. Data for simulations and testing were collected on 3T Siemens scanners.

\section{S\lowercase{imulations}}
Recordings containing scanner noise only and speech only were used for simulations. The formulae used for quantifying initial signal to noise ratio (SNR) and improved SNR (ISNR) are listed in Table~\ref{tab:SNRform}.

\begin{table}
\caption{\label{tab:SNRform}Formulae for quantifying simulations}
\begin{ruledtabular}
\begin{tabular}{lccc}
Value	& Formula\\
\hline
Noise suppresion(NS) & $20\log_{10}\frac{||\mathbf{g}-\mathbf{\hat{g}}||_2}{||\mathbf{g}||_2}$\\
Signal to noise ratio (SNR) & $20\log_{10}\frac{||\mathbf{v}||_2}{||\mathbf{g}||_2}$\\
Improvement in SNR (ISNR) & $20\log_{10}\frac{||\mathbf{g}||_2}{||\mathbf{v}-\mathbf{\hat{v}}||_2}$\\
\end{tabular}
\end{ruledtabular}
\end{table}

\begin{figure}
\includegraphics[width=.9\columnwidth]{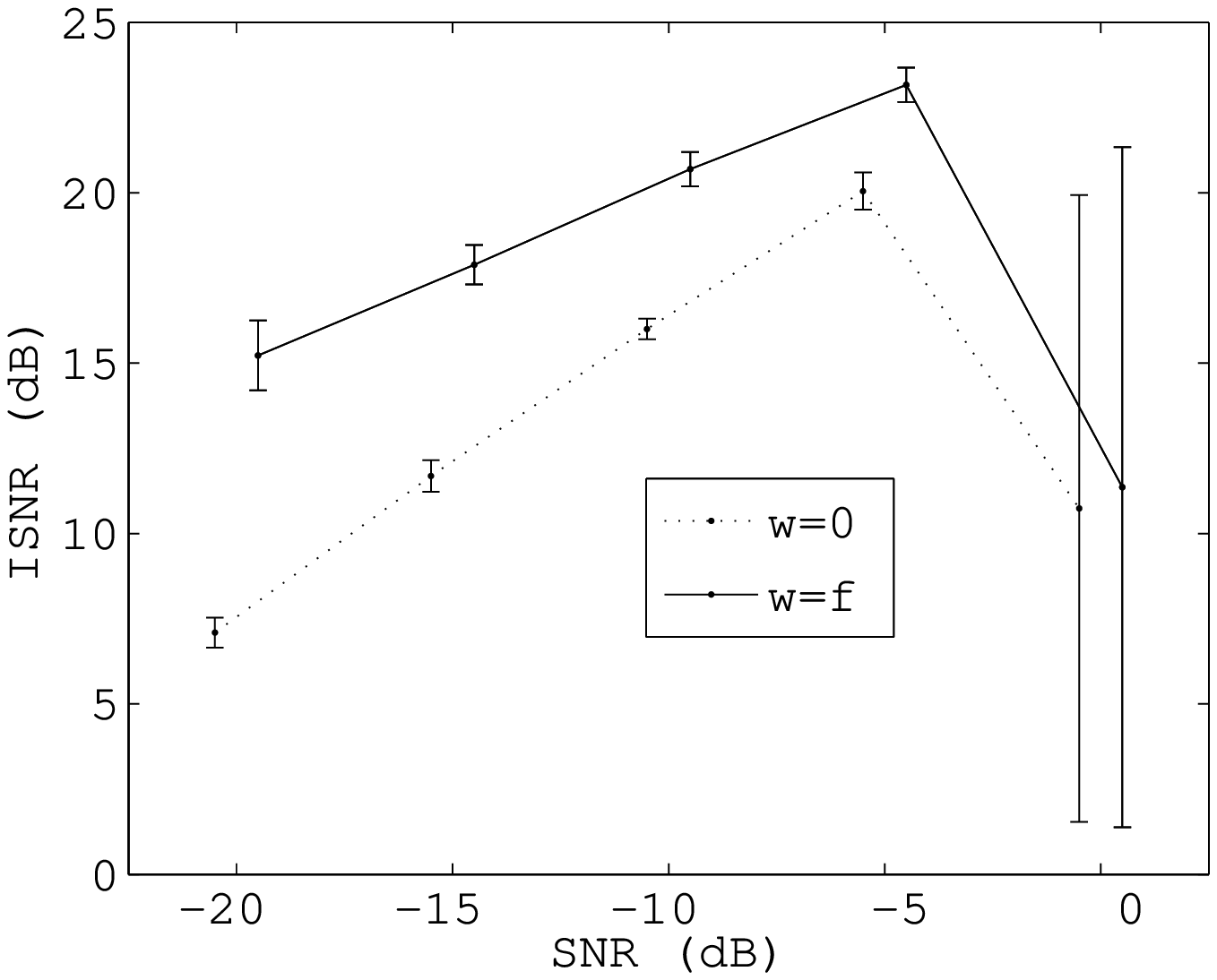}
\caption{Improvement in signal-to-noise ratio (ISNR) shown for two weighting functions over a range of SNRs.}
\label{fig:isnrwt}
\end{figure}

\begin{figure*}
\includegraphics[width=\textwidth]{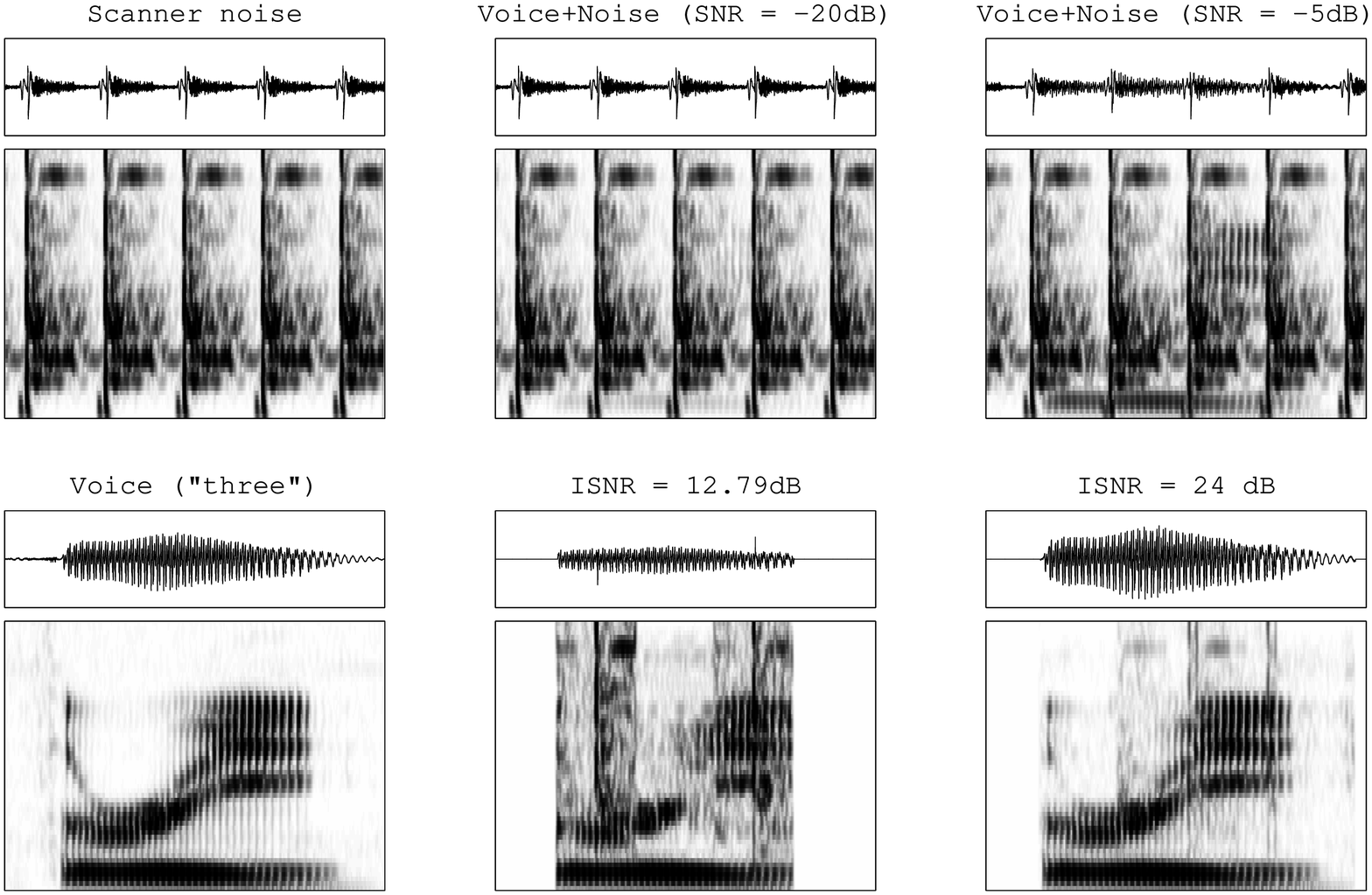}
\caption{The left panel shows the time-domain waveforms and corresponding spectrograms of the speech (the utterance ``three'') and the scanner noise signals. The middle and the right panels of the top row show the synthesized noise-corrupted signals (top: SNR = -20dB and bottom SNR = -5dB). The middle and the right panels of the bottom row show the noise reduced waveforms, spectrograms and the improvement in SNR (ISNR). Spectrograms have a frequency range from 0-5KHz.}
\label{fig:acoustresults}
\end{figure*}

\subsection{Influence of weighting functions: $\mathbf{w}$}
Frequency-domain weighting functions $\mathbf{w}$ differentially affect the amount of noise suppression (NS) achieved on noise only sequences. A weighting function $\mathbf{w}=0$ corresponds to time domain subtraction, while the function $\mathbf{w}=1$ is equivalent to increasing the suppression parameter $\alpha$. Not surprisingly, for recovery of speech signals, the most effective weighting function was one that suppressed higher frequency content more than low frequency content. To quantify the difference, ISNR were measured by comparing the SNR before and after noise suppression of synthesized data.

Single word speech utterances (e.g., Mm.1), $v(t)$, were added to scanner noise $g(t)$ with different signal to noise ratios (SNR) and relative phase (e.g., Mm. 2 and Mm. 3). Two different weighting functions were used: (1) $\mathbf{w} = 0$; and (2) $\mathbf{w}(f) = |f|$. Improvement in signal to noise ratio (ISNR) was calculated over the utterance only. Figure~\ref{fig:isnrwt} shows ISNR as function of original SNR and the different weighting functions. The variance is computed over different values of $\alpha$ and $\theta_{RMS}$.

Mm. 1. Recorded speech used in simulation. This is a file of type ``wav''. 
Mm. 2. Synthesized mixture (SNR = -20 dB) used in simulation. This is a file of type ``wav''.
Mm. 3. Synthesized mixture (SNR = -5 dB) used in simulation. This is a file of type ``wav''.

In general, a higher original SNR leads to a higher ISNR. However, at SNRs close to 0 dB, the speech signal disrupts the pulse-train like nature of the scanner noise and violates the $|g(t)|>>|v(t)|$ assumption. This results in greater misalignment in step 2 of the algorithm and therefore reduces the amount of improvement in SNR. Figure~\ref{fig:acoustresults} shows the acoustic results from two of the simulations, one at -20dB SNR (Mm. 4) and the other at -5dB SNR (Mm. 5). Results of processing actual recordings made in the scanner are available as Mm. 6 and Mm. 7.

Mm. 4. Cleaned recording (-20db SNR) processed using the algorithm. This is a file of type ``wav''.
Mm. 5. Cleaned recording (-5dB SNR) processed using the algorithm. This is a file of type ``wav''.
Mm. 6. Recorded speech in scanner.  This is a file of type ``wav''.
Mm. 7. Cleaned recording processed using the algorithm. This is a file of type ``wav''.

\section{C\lowercase{onclusion}}
We have presented an adaptive, online algorithm that can be used to suppress scanner noise and thereby provide an effective channel of communication between the subject in the scanner and the operator. It also allows acquisition of verbal responses in fMRI studies. The algorithm combines time-domain and frequency-domain techniques for noise reduction and can be easily implemented on a computer or a DSP board for specialized operation. The numerical description of the noise cancellation process provides a generalized framework that can be easily extended. The parameters can be easily optimized for different pulse sequences and other similar repetitive signals. 

\section*{A\lowercase{cknowledgements}}
The study was partially supported by NIH grants R01 DC02852 (Frank Guenther, PI) and R01 DC01925 (Joseph Perkell, PI). We would like to thank Jay Bohland, Alfonso Nieto-Castanon, Kim Lumbard and Lawrence Wald for help with scanning and comments on the manuscript. Use of the Athinoula A. Martinos Center for Biomedical Imaging facilities was aided by support from National Center for Research Resources grant P41RR14075 and the MIND Institute.

%The following command formats the BiBTeX-generated bibliography by reading in the .bbl file.
%When preparing a TeX document for submission to the JASA, you must paste in the contents
%of that file in place of this command: the Journal requires submission of a single .tex file.
%\bibliography{NS01_jasa}%

\begin{thebibliography}{4}
\newcommand{\enquote}[1]{``#1''}
\expandafter\ifx\csname natexlab\endcsname\relax\def\natexlab#1{#1}\fi
\expandafter\ifx\csname url\endcsname\relax
  \def\url#1{\texttt{#1}}\fi
\expandafter\ifx\csname urlprefix\endcsname\relax\def\urlprefix{URL }\fi
\providecommand{\bibinfo}[2]{#2}
\providecommand{\noopsort}[1]{}
\providecommand{\switchargs}[2]{#2#1}

\bibitem[{Cusack \emph{et~al.}(2005)Cusack, Cumming, Bor, Norris, and
  Lyzenga}]{cusa.automated.05}
\bibinfo{author}{Cusack, R.}, \bibinfo{author}{Cumming, N.},
  \bibinfo{author}{Bor, D.}, \bibinfo{author}{Norris, D.}, and
  \bibinfo{author}{Lyzenga, J.} (\textbf{\bibinfo{year}{2005}}).
  \enquote{\bibinfo{title}{Automated post-hoc noise cancellation tool for audio
  recordings acquired in an mri scanner.}}, \bibinfo{journal}{Hum Brain Mapp}
  \textbf{\bibinfo{volume}{24}}, \bibinfo{pages}{299--304},
  \urlprefix\url{http://dx.doi.org/10.1002/hbm.20085}.

\bibitem[{Jung \emph{et~al.}(2005)Jung, Prasad, Qin, and
  Anderson}]{jung.extraction.05}
\bibinfo{author}{Jung, K.-J.}, \bibinfo{author}{Prasad, P.},
  \bibinfo{author}{Qin, Y.}, and \bibinfo{author}{Anderson, J.~R.}
  (\textbf{\bibinfo{year}{2005}}). \enquote{\bibinfo{title}{Extraction of overt
  verbal response from the acoustic noise in a functional magnetic resonance
  imaging scan by use of segmented active noise cancellation.}},
  \bibinfo{journal}{Magn Reson Med} \textbf{\bibinfo{volume}{53}},
  \bibinfo{pages}{739--744},
  \urlprefix\url{http://dx.doi.org/10.1002/mrm.20398}.

\bibitem[{Nelles \emph{et~al.}(2003)Nelles, Lugar, Coalson, Miezin, Petersen,
  and Schlaggar}]{nell.automated.03}
\bibinfo{author}{Nelles, J.~L.}, \bibinfo{author}{Lugar, H.~M.},
  \bibinfo{author}{Coalson, R.~S.}, \bibinfo{author}{Miezin, F.~M.},
  \bibinfo{author}{Petersen, S.~E.}, and \bibinfo{author}{Schlaggar, B.~L.}
  (\textbf{\bibinfo{year}{2003}}). \enquote{\bibinfo{title}{Automated method
  for extracting response latencies of subject vocalizations in event-related
  fmri experiments.}}, \bibinfo{journal}{Neuroimage}
  \textbf{\bibinfo{volume}{20}}, \bibinfo{pages}{1865--1871}.

\bibitem[{Ravicz \emph{et~al.}(2000)Ravicz, Melcher, and
  Kiang}]{ravi.acoustic.00}
\bibinfo{author}{Ravicz, M.~E.}, \bibinfo{author}{Melcher, J.~R.}, and
  \bibinfo{author}{Kiang, N.~Y.} (\textbf{\bibinfo{year}{2000}}).
  \enquote{\bibinfo{title}{Acoustic noise during functional magnetic resonance
  imaging.}}, \bibinfo{journal}{J Acoust Soc Am}
  \textbf{\bibinfo{volume}{108}}, \bibinfo{pages}{1683--1696}.

\end{thebibliography}

\end{document}